\documentclass[traditabstract]{aa} 
\usepackage{graphicx}
\usepackage{txfonts}
\begin{document}
   \title{Pulsed Gamma-Ray-Burst Afterglows}


   \author{J. Middleditch
          \inst{1}
          }

   \institute{Los Alamos National Laboratory, Los Alamos,
              NM 87545\\
              \email{jon@lanl.gov}
             }


 
  \abstract
{The bipolarity of Supernova 1987A can be understood
in terms of its very early light curve as observed from the CTIO 
0.4-m telescope, as well as the IUE FES, and the slightly later 
speckle observations of the ``Mystery Spot'' by two groups.  These 
observations imply a highly directional beam of light and jet of 
particles, with initial collimation factors in excess of 10$^4$, 
velocities in excess of 0.95 c, as an impulsive event involving 
up to 10$^{-5}$ M$_{\bigodot}$, which interacts with circumstellar
material.  The jet and beam coincide with the 194$^{\circ}$ angle 
of the bipolarity on the sky, and are oriented at 75$^{\circ}$ 
to the line of sight to the Earth.  By day 30 the 
collimation of the jet decreases, and its velocity declines to 
$\sim$0.5 c.  These observations and the resulting kinematic
solution can be understood in terms of pulsar emission from 
polarization currents, induced by the periodically modulated 
electromagnetic field beyond the pulsar light cylinder,
which are thus modulated at up to many times the speed of light.
With plasma available at many times the light cylinder radius,
as would be the case for a spinning neutron star formed at the
center of its progenitor, pulsed emission is directed close to
the rotation axis, eviscerating this progenitor, and continuing
for months to years, until very little circumpulsar material remains.
There is no reason to suggest that this evisceration mechanism is not 
universally applicable to all SNe with gaseous remnants remaining.
Calculations of this mechanism are orders of magnitude more difficult
than previously imagined for any pulsar interaction with its remnant
progenitor.  This model provides a candidate for the central engine 
of the gamma-ray burst (GRB) mechanism, both long and short,
and predicts that GRB afterglows are the {\it pulsed} optical/near 
infrared emission associated with these newly-born neutron stars.  
It also provides a mechanism to accelerate electrons and positrons to 
ultrarelativistic energies,
possibly accounting for the results from PAMELA and ATIC, and
the WMAP haze.  It is also possible, within the context of this
model, that the prompt emission from the gamma-ray burst itself
may result from a white dwarf near Chandrasekhar mass rotating with
its minimum period near 2 s, rather than from the more rapidly
rotating neutron star formed from its subsequent collapse.
We note that the bipolarity, enforced on early SN remnants by their 
embedded pulsars, i.e., very fast axial ejection features within 
expanding toroids, may complicate their utility, as standard
candles, to cosmological interpretation.}



   \maketitle
%

\section{Introduction}
\label{sec:intro}

\begin{figure}
   \centering
   \includegraphics{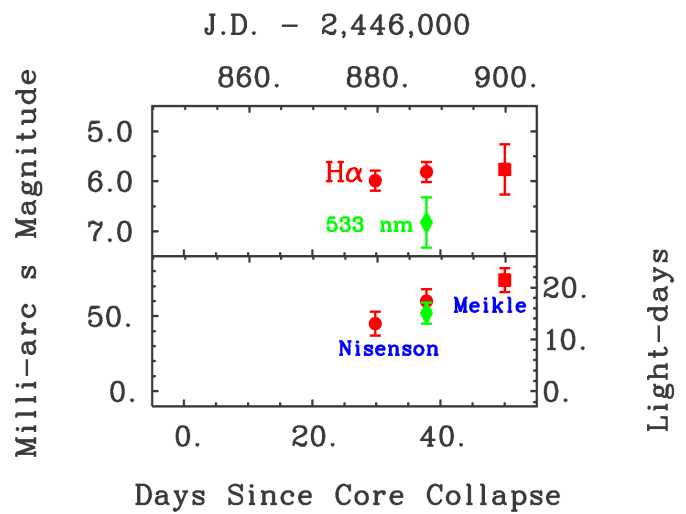}
   \caption{Measurements of displacement (lower) and observed
   magnitude (upper) of the ``Mystery Spot''from SN 1987A,
   at H$_{\alpha}$ and 533 nm, vs time, from Meikle et al.~(1987),
   Nisenson et al.~(1987), and Nisenson and Papaliolios (1999).
	       }
              \label{magdist}%
    \end{figure}

Supernova 1987A has provided astronomers with a wealth of data,
some of which has not even now, 22 years after the event, been
satisfactorily accounted for by any model.
One of the most remarkable features
in the early study of SN 1987A was the ``Mystery Spot'',
with a thermal energy of 10$^{49}$ ergs, observed 50 days after
the core-collapse event (Meikle et al.~1987; Nisenson et al.~1987;
Nisenson \& Papaliolios 1999), and separated from the SN photosphere
``proper'' by 0.060$\pm$0.008 arc s at day 38 (Fig.~\ref{magdist}), with about 3\%
of this energy eventually radiated in the optical band.
The possibility that the enormous energy implied for the Mystery Spot might
somehow link it to gamma-ray bursts (GRBs) attracted little serious
consideration at the time, or even since, beyond a very astute few
(\cite{Rees87}; \cite{PN87}; \cite{Cen99}).  The Mystery Spot was 
also observed at separations
of 0.045$\pm$0.008 arc s on day 30, and 0.074$\pm$0.008 arc s on day
50, but always at an angle of 194$^{\circ}$, consistent with the southern
(and approaching) extension of the bipolarity (\cite{Wa02}).  The Mystery 
Spot offsets from SN 1987A imply a minimum projected separation of $\sim$10 
light-days ($\ell$t-d).

There is a also wealth of photometric and spectroscopic data from even 
earlier stages of SN 1987A, in particular photometry data from the
Cerro Tololo Inter-American Observatory (CTIO) 0.4-m telescope (\cite{HS90}), 
and the Fine Error Sensor (FES) of the International Ultraviolet Explorer 
(IUE -- Wamsteker et al.~1987), and spectroscopic data from 
Danziger et al.~(1987), and Menzies et al.~(1987), among others. 
Short timescale structure ($\le$1 d) in this data, following finite
delays ($\sim$10 d) after SN 1987A core-collapse, implies at least
one beam of light and jet particles which are highly collimated
($>$10$^4$), interacting with circumstellar material.

GRBs, particularly long, soft GRBs ($\ell$GRBs), appear to be the most
luminous objects in the Universe, occurring at the SN rate of one per
second (assuming a collimation factor near 10$^5$) yet we still know
very little about them (see, e.g., M\'esz\'aros 2006 and references
therein).  Although some have been found to be associated with SNe,
others, mostly those with slightly harder spectra and lasting only
$\sim$1 second, (sGRBs), produce {\it only} ``afterglows'' (if
that), sometimes extending down to radio wavelengths.  A large number
of models have been put forth to explain GRBs, including neutron
star-neutron star mergers
for sGRBs, and exotic objects such as ``collapsars'' (\cite{MW99}) for
$\ell$GRBs.  The prime physical motivation for these is the enormous
energy of up to 10$^{54}$ ergs implied for an isotropic source.
However, given that the data from SN 1987A presented herein support a
beam/jet collimation factor $>$10$^4$ in producing its
early light curve by interaction with more-or-less stationary
circumstellar material
(see below), there may be no need for such a high energy.

This work argues that polarization currents,
induced beyond the light cylinders of, and
by the rotating magnetic fields from,
newly-formed pulsars embedded within their stellar remnants
(\cite{KPR74}; \cite{BR01}), can account for the bipolarity of SN 1987A
(Ardavan 1994,8; \cite{Ar08}).
This model of emission from superluminally induced polarization currents 
(SLIP) provides a mechanism for generating a pulsed beam on the 
surface of a cone, whose half angle (and angle from the pulsar 
axis of rotation) is given by,
\begin{equation}
\theta_V = \sin^{-1} {\rm c}/v,
\end{equation}
for astronomical distances.  Here c is the speed of light, 
and $v>{\rm c}$ is the speed at which the
polarization currents are updated, i.e., $v = \omega R$,
where $\omega$ is the pulsar rotation frequency, in radians s$^{-1}$,
$R>R_{LC}$ is the distance of the polarization current from the pulsar,
projected onto the rotational equatorial plane,
and $R_{LC}$ is the light cylinder radius ($\omega R_{LC} = {\rm c}$).
The power emitted rises steeply with $v$ (\cite{Ar04}), and this beam,
in turn acting as a phased array, accelerates matter into a conical jet, 
centered about the axis of rotation.

   \begin{figure}[h!]
   \centering
   \includegraphics{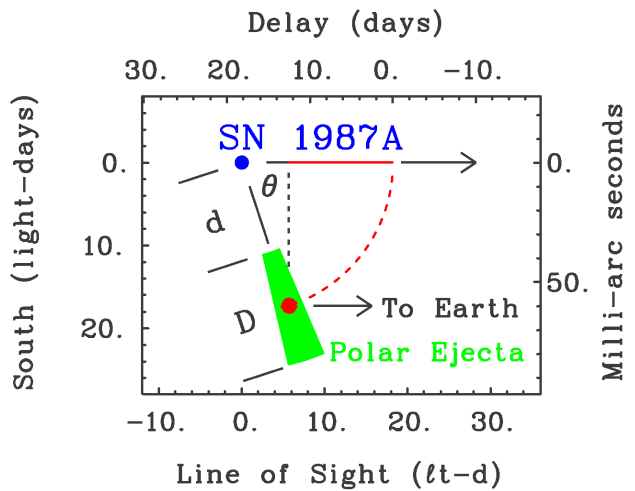}
   \caption{
   The geometry of the ``Mystery Spot,'' (red dot)
   polar ejecta, associated beam/jet, and direct line of sight
   from SN 1987A to the Earth.
	       }
              \label{fgeom}%
    \end{figure}

In the rest of this paper, Sect.~\ref{early} includes a quantitative 
discussion of the SN 1987A early luminosity history and motivations 
for why a later, quasi-steady, less collimated, as well as a prompt, 
highly collimated, injection event, is also needed.  Then Sect.~\ref{geom} 
estimates the kinematics and a working geometry for the 1987A 
beam/jet and Mystery Spot.  We explore the implications of the
kinetics and observations on the process which gave rise to SN 1987A.
Section \ref{link} relates the SN 1987A beam/jet process to GRBs and 
germinates the idea that they and their afterglows are highly pulsed, 
while Sect.~\ref{Ia/c} relates the process to Type Ia/c SNe.  Section 
\ref{TypeII} extends the discussion of implications of the SN 1987A 
process to Type II SNe, the importance of plasma to weakly magnetized 
pulsars, and its role in the history of observations of SN 1987A, as 
well as the consequences of the motion of the Mystery Spot, and of 
the high kinetic energy of particles in the SN 1987A jet(s).  Finally, 
Section \ref{conc} concludes.

\section{The Early Luminosity History of SN 1987A}
\label{early}

\begin{figure}
   \centering
    \includegraphics{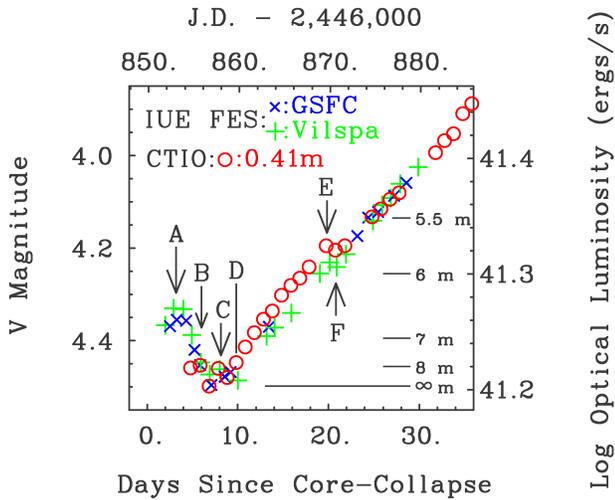}  
   \caption{
The very early luminosity history of SN 1987A as observed
with the Fine Error Sensor of IUE and the 0.41-m at CTIO.  Data
points taken at Goddard Space Flight Center by Sonneborn \& Kirshner,
and at the Villafranca Station in Madrid, are marked (see
Sect.~\ref{early}).
	       }
              \label{fearly}%
    \end{figure}

Table 1 gives an event history for SN 1987A and its
progenitor system. 
An approximate geometry for SN 1987A and a Mystery Spot located 
within circumstellar material (in this case, polar ejecta), is
given in Fig.~\ref{fgeom}, while the early luminosity 
histories of SN 1987A from CTIO, and the Fine Error Sensor 
(FES) of the IUE, are both plotted in
Fig.~\ref{fearly}.\footnote{The CTIO V band
center occurs at 5,500 \AA, as opposed to 5,100
\AA ~for the FES, and in consequence, the FES magnitudes have
been diminished by 0.075 in Fig.~\ref{fearly} to account for the resulting
luminosity offset, and the CTIO times (\cite{HS90}) are too early by
1 day, and have been corrected in this work.}

\begin{table}[ht!]
\caption{SN 1987A Event Log}             
\label{table:1}      
\centering                          
\begin{tabular}{l l l }        
\hline\hline                 
Time      &&Event \\    
 t                 \\    
\hline                        
-20,000 years &&Rings formed \\ 
$\sim$(-2,000?) years &&Polar, or near-polar ejection \\
 0 (= UT 1987, Feb. 23.316) && Core-collapse of SN 1987A \\
 0$<$t$<$2 days  &&UV flash from SN 1987A  \\
 2$<$t$<$4 days  &&Emergence of luminous jet \\
 4$<$t$<$7 days  &&Cooling/spreading of jet \\
 7.8 days    &&UV flash hits polar ejecta\\
 8.26 days   &&Jet impacts polar ejecta (PE) \\
 19.8 days     &&Pulsations clear through PE \\
 20.8 days     &&Jet particles clear through PE \\
 30 days     &&``Mystery spot'' at 45 mas  \\
 38 days     &&``Mystery spot'' at 60 mas  \\
 50 days     &&``Mystery spot'' at 74 mas  \\
\hline                                   
\end{tabular}
\end{table}

\begin{figure}[!h]
   \centering
    \includegraphics{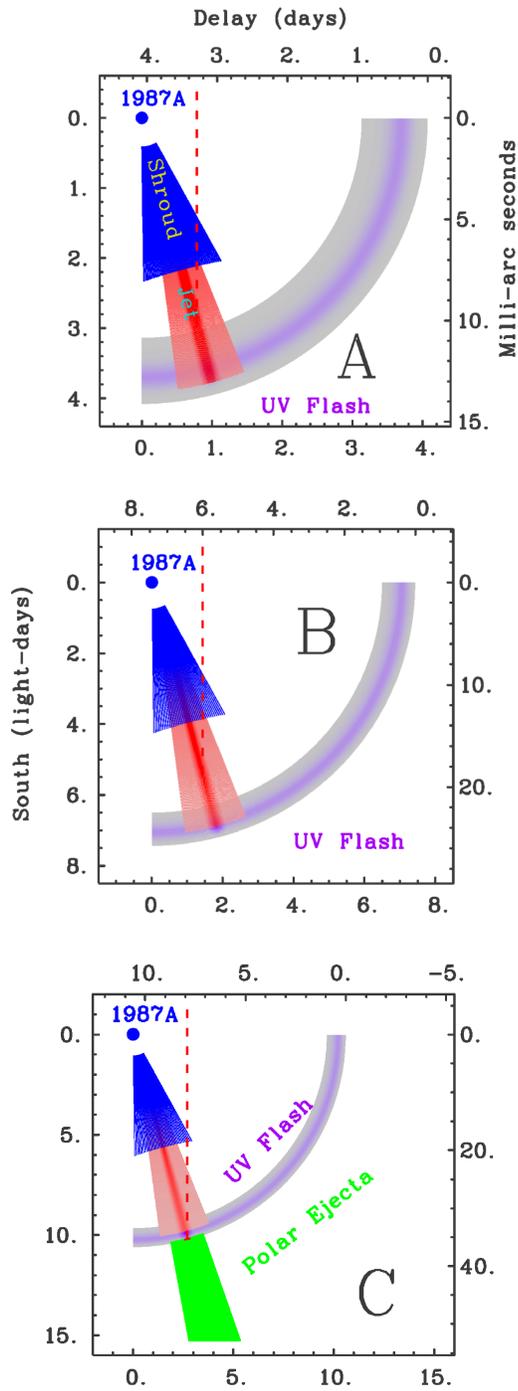}  
   \caption{
The geometry of the 1987A glowing beam/jet, initially opaque shroud, 
and UV flash (which may have an enhanced beam of its own in the jet 
direction -- here 75$^{\circ}$, down and to the right).  
The maximum velocity of the jet/shroud is 0.95/0.55 c.
The dashed line to the upper scale flags the center of the emerging 
jet at day 3.3 (top -- A), and day 6 (middle -- B), and the UV flash 
hitting the polar ejecta at day 7.8 (bottom -- C).
	       }
              \label{ABC}%
    \end{figure}

Following the drop from the initial flash, the luminosity rises
again to a maximum (`A' in Fig.~\ref{fearly} and Fig.~\ref{ABC}, 
top) of magnitude 4.35 at day 3.0, roughly corresponding to 
2.7$\times$10$^{41}$ ergs s$^{-1}$
and interpretable as a luminous jet emerging from cooler, roughly
cylindrical outer layers which initially shrouded it.  This declines 
to magnitude 4.48 around day 7.0 (`B', Fig.~\ref{ABC}, middle), interpretable 
as free-free cooling, or the loss of the ability to cool, as the
jet becomes more diffuse.  The next observable event should be
the scatting/reprocessing of the initial UV flash in the polar 
ejecta at day $\sim$8, and indeed `C' (Fig.~\ref{ABC}, bottom)
shows
$\sim$2$\times$10$^{39}$ ergs s$^{-1}$ in the optical for a day at
day 7.8, and a decline {\it consistent with the flash} after that,
indicating that no significant smearing over time had occurred in 
this interaction.  The 8 day delay to this first event implies a
collimation factor $>$10$^4$ for this part of the UV flash.

\begin{figure}[!h]
   \centering
    \includegraphics{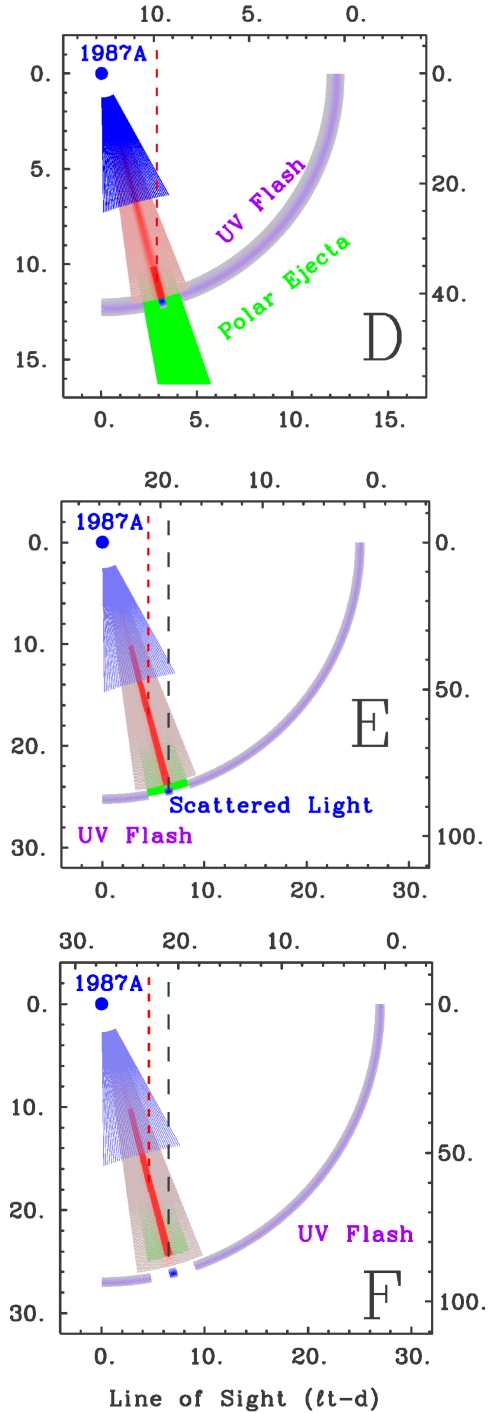}  
   \caption{
(Top -- D) The intense center ($\sim$1$^{\circ}$) of the jet begins to 
produce light as it penetrates into the polar ejecta, producing 
the jump in luminosity at day 9.8.
(Middle -- E) Particles in the jet continue to impact the polar ejecta 
(mostly hidden), extending the ramp in luminosity visible in 
Fig.~\ref{fearly} near day 20 (left dashed line to the top scale).  
(Right dashed line) Light from the filtered UV flash scatters in 
clumpy polar ejecta near its outer boundary.  (Bottom -- F) The fastest 
jet particles have cleared the end of the polar ejecta. 
	       }
              \label{DEF}%
    \end{figure}

A linear ramp in luminosity, visible by day 9.8 ('D' in Fig.~\ref{fearly}
and Fig.~\ref{DEF}, top), 
indicates particles from 
the jet penetrating into the polar ejecta, with the fastest traveling at
$>$0.8 c, {\it and}, because of the sudden rise, a particle collimation 
factor $>$10$^4$ (see further 
below).  Back-extrapolation of the three CTIO points just after day 8
intersects the day 7 minimum near day 8.26, which would indicate
that particles exist in this jet with velocities up to 0.95 c,
and even higher if the true minimum flux is lower than the points
at magnitude 4.48 (1.6$\times$10$^{41}$ erg s$^{-1}$) near day 7.

The ramp continues until after day 20, when a decrement  
of $\sim$5$\times$10$^{39}$ ergs s$^{-1}$ appears in both
data sets just after day 20 (`F' in Fig.~\ref{fearly}, and 
Fig.~\ref{DEF}, bottom).  
The CTIO point just before 
the decrement can be used as a rough upper limit for the Mystery Spot 
luminosity, and corresponds to an
excess above the minimum (near day 7.0) of 5$\times$10$^{40}$ ergs
s$^{-1}$, or magnitude 5.8, the same as that observed in H$\alpha$
for the  at days 30, 38, and 50, about 23\% of the total optical flux 
of 2.1$\times$10$^{41}$ ergs s$^{-1}$ at that time.  

\begin{figure}
   \centering
    \includegraphics{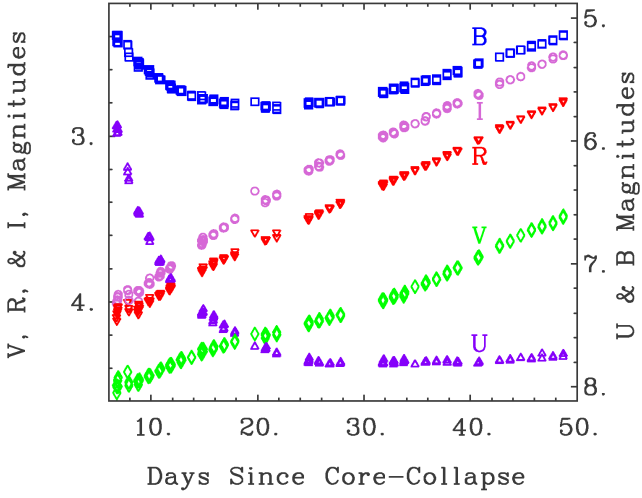}  
   \caption{
The U, B, V, R, and I points from the CTIO 0.41-m from
days 6 to 50 (see Sect.~\ref{early}).
	       }
              \label{ubvri}%
    \end{figure}

This decrement is preceded by a {\it spike} ('E' in Fig.~\ref{fearly},
Fig.~\ref{DEF}, middle, and Fig.~\ref{ubvri}, day 19.8) of up to 
10$^{40}$ ergs s$^{-1}$ in the CTIO data, with the unusual colors 
of B, R, and I, in ascending order, very close to the B and I bands 
speculated for the 2.14 ms signature observed from 1987A by Middleditch 
et al.~(2000a,b -- hereafter M00a,b), with an H$_{\alpha}$ enhancement.
Spectra taken by Danziger et al.~(1987), on UT 1987, March 15.08 
(day 19.76), and Menzies et al.~(1987), on March 14.820 (day 19.504),
support these flux enhancements, including the H$_{\alpha}$.  
The timing of this event, one 
day prior to the decrement, suggests that it is due to a photon stream, 
stripped of its UV component by absorption (the CTIO U point at day 19.8
was low, consistent with this interpretation), scattering into other
directions, including the line of sight to the Earth, by what might
have been a clumpy end to the circumstellar material.  Pulsations
were not detected (R.~N.~Manchester, private communication), because 
of the oblique view, and the dimensions of the beam ($\sim$1 $\ell$t-d).


For the geometry
derived in Sect.~\ref{geom} below, the one day delay implies 
at least the same maximum jet velocity (0.95 c), supporting this 
interpretation, and giving us a rough isotropic lower limit estimate 
of the strength of the pulsations.  Spectra taken just before day 5,
showing an enhancement for wavelengths below 5000 $\AA$, explain the 
discrepancy between the CTIO and FES points at that time 
(Fig.~\ref{fearly}).

In spite of the coincidence between the end magnitude of the linear
ramp and that of the Mystery Spot, the two are probably not the same
effect, as the offset of the Mystery Spot from SN 1987A was only 0.045 arc
s even 10 days later at day 30 (Fig.~\ref{spot}), a location barely 
beyond where the
ramp began, as is shown below, and there is no sign of elongation
toward 1987A proper in Fig.~1 of Nisenson and Papaliolios (1999) or
Fig.~2 of Nisenson et al.~(1987).  The Mystery Spot may develop as a plume
within the polar ejecta, pushed by a less collimated, 0.5 c pulsar
wind, perhaps not unlike that observed from the Crab pulsar
(\cite{He02}), after the passage of the initial, very fast,
very collimated component of the jet.  A beam only 1 $\ell$t-d across
at $\sim$10 $\ell$t-d translates into plasma at $\sim$20 $R_{LC}$.
Alternately, the early light curve might be due to shallow penetration 
of a precessing jet into a varying entry point into the polar ejecta.  
However, the high density required to limit jet penetration
comes with a higher opacity which would make the linear ramp hard
to produce in this, the inner boundary of the approaching axial feature,
and the requirement for a 0.5 c mean motion of the Mystery Spot between 
days 30 and 38, slowing to 0.35 c between days 38 and 50 (ostensibly due
to swept up matter), would also be difficult to account for in these 
circumstances.  On the other
hand, the polar ejecta density can not be so low as to allow more
than 1 $\ell$t-d penetration by the enhanced UV flash, or the
drop in luminosity from day 7.8 to day 8.8 would not be as big.  If the
jet penetration is deep, precession and/or changes in the plasma density
beyond the pulsar light cylinder (Eqn.~1), may make its initial track,
within the polar ejecta, helical, and this may assist the $\sim$0.5 c
wind in the creation of the plume which forms the Mystery Spot within 
three weeks of the initial jet penetration.

We will assume that the optical flux from the interaction between
jet particles and the polar ejecta will not be significantly occulted 
in the ejecta itself in the direction to the Earth, otherwise again,
the linear ramp would be difficult to produce.  As we will find
below that the axis of the SN 1987A bipolarity is $\sim$30$^{\circ}$
from the normal to the ring planes, the reason remaining a mystery
even today, this is not necessarily a given.  Proceeding nevertheless:
by scaling homologously inward a factor of 10 from the equatorial ring
density of 10$^4$ cm$^{-3}$, we arrive at a polar ejecta density
estimate of $\sim$10$^7$ cm$^{-3}$ -- sufficient to stop the UV flash
from penetrating $>$1 $\ell$t-d.  

Assuming a polar ejecta depth, D, of $\sim$10
$\ell$t-d, or 2.6$\times$10$^{16}$ cm, gives a total column of
2.6$\times$10$^{23}$ cm$^{-2}$, enough to warrant some concern.
However, only a fraction of the protons in the jet will
scatter through the entire depth of the polar ejecta (the slight
concave downward departure from linearity most apparent in
the CTIO data, between days 9 and 20, may reflect this loss,
and/or the density in the polar ejecta may decrease with distance).
In addition, we will find that the angle, $\theta$, from our line of
sight to the SN 1987A beam/jet, will be large in the self-consistent
solution, justifying our assumption of visibility for
the luminous column within the polar ejecta between days 9 and 20.

\section{The Geometry and Kinematics of the Beam/Jet}
\label{geom}

Using the constraints shown in Figs.~\ref{magdist} and \ref{fearly}, 
we can solve for the
three geometric variables, $d$, $D$, and $\theta$, diagrammed in
Fig.~\ref{fgeom},
and the maximum velocity of the particles in the jet, $\beta$.
First the UV flash hits the beginning of the polar
ejecta at day 7.8:
\begin{equation}
d(1-cos \theta)  = {\rm c}t_0 == 7.8~{\rm \ell t-d},
\end{equation}
where $d$ is the distance to the beginning of the polar ejecta,
$\theta$ is the angle from the line of sight to the Earth
to the beam/jet/polar ejecta direction, and c is the speed of
light.

From Fig.~\ref{fearly} we also have the jet particles well into 
the polar ejecta by day 9.8.  Extrapolating backward per above, we have 
the fastest beam particles hitting the polar ejecta at day 8.26:
\begin{equation}
d(1/\beta-cos \theta ) = {\rm c}t_1 == 8.26~{\rm \ell t-d}.
\end{equation}
Next, we have the projected offset of 0.060 arc s for the Mystery Spot,
measured at day 29.8 by Nisenson et al.~(1987) and refined
by Nisenson and Papaliolios (1999).  This is more difficult
to pin down relative to its position radially through the polar
ejecta,
so we assume it's some fraction, $\alpha$, of the way through
the polar ejecta depth, $D$, and hope for a self-consistent solution:
\begin{equation}
(d+\alpha D) \sin \theta = {\rm c}t_2 == 17.3~{\rm \ell t-d},
\end{equation}
using 50 kpc for the distance to SN 1987A.
Finally, we have the decrement in the light curve at day 20,
shown in Fig.~\ref{fearly}, which we will interpret as the fastest
``substantial'' bunch of particles in the jet breaking
through the end of the polar ejecta:
\begin{equation}
(d + D)(1/\beta - cos \theta ) = {\rm c}t_3 == 20~{\rm \ell t-d}.
\end{equation}

The solution to Eqns.~1-4 gives a constant ratio for $D$ to $d$,
independent of $\alpha$:
\begin{equation}
d  = D t_1/(t_3 - t_1),~{\rm or}~(d + D) = d~t_3 / t_1,
\end{equation}
while $\theta$ is given by:
\begin{equation}
\theta  = 2 \tan^{-1} \{{t_0 \over t_2}(\alpha({t_3 \over t_1}-1)+1)\}.
\end{equation}
The parameters, $d$, $D+d$, and $\theta$ are plotted against
$\beta$ in Fig.~\ref{soln} for 0.3 $\le$ $\alpha$ $\le$ 0.7,
along with the maximum $d$ and minimum $D + d$ implied by the
three measurements of the
Mystery Spot angular separation at days 30, 38, and 50.
Figure \ref{soln} shows that the polar ejecta, at the very least, 
must start by 14 $\ell$t-d or closer, and extend to 22 $\ell$t-d or farther,
consistent with our early 10 $\ell$t-d estimates for $d$ and $D$,
as is the high value of $\theta$ ($65^{\circ}<\theta<85^{\circ}$),
which also means that the axis of bipolarity is $\sim$30$^{\circ}$ 
from the normal to the ring planes (\cite{Su05}).

   \begin{figure}
   \centering
   \includegraphics{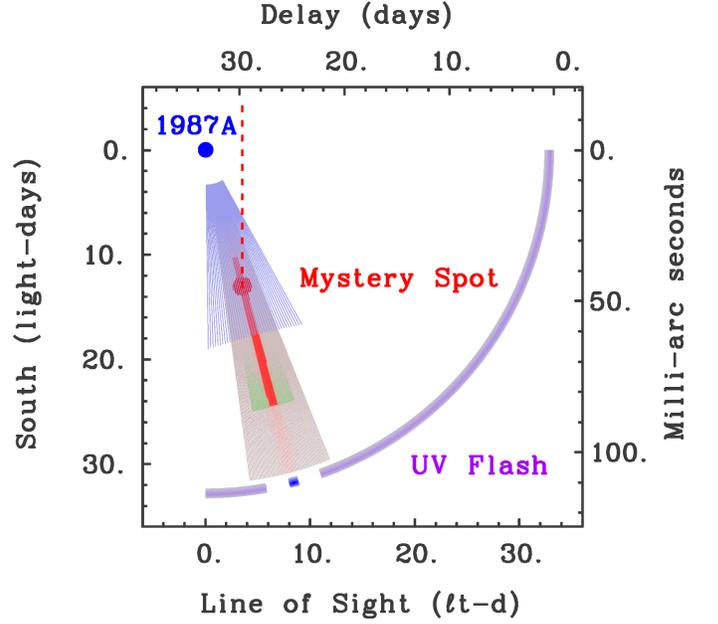}
   \caption{ 
   The relation of the Mystery Spot, near day 30, to a jet,
   a thinning shroud, and a UV Flash, when its offset from 
   SN 1987A was 0.045 arc s.
	       }
              \label{spot}%
    \end{figure}

   \begin{figure}
   \centering
   \includegraphics{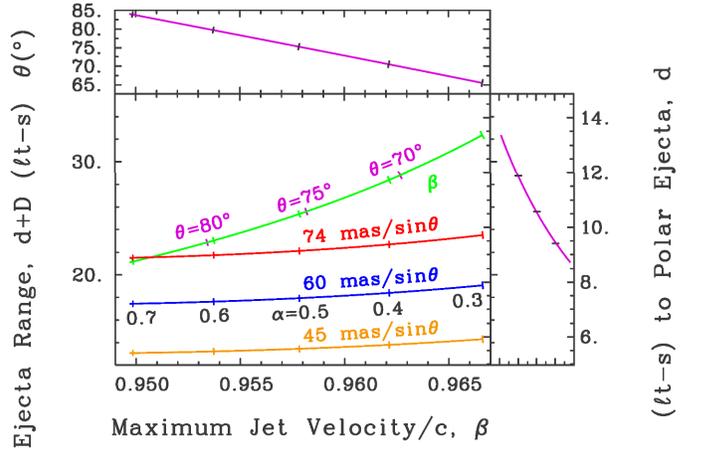}
   \caption{
   The solution values for Eqns.~1-4, (Big frame horizontal)
   The maximum jet velocity, $\beta$.  (Left vertical) The maximum
   range of the
   polar ejecta.  (Right vertical axis) The distance from the pulsar to the
   beginning of the polar ejecta.  The line with the steepest slope
   matches $\beta$ (bottom) to $D + d$ (left), or $d$ (right), and
   three values for $\theta$ are marked.  The three other lines
   with moderate slopes constrain the minimum of $D + d$ (right
   end of 74 mas curve), and the maximum of $d$ (left end of the
   45 mas curve and also read on the left vertical axis) from offset
   measurements of the Mystery Spot, which is assumed to be
   a jet-driven plume {\it within} the polar ejecta.  (Top
   frame) Theta as a function of beta.  (Right frame) Theta
   (horizontal axis identical to top frame vertical axis) versus
   $D + d$ (left vertical), or $d$ (right vertical).
	       }
              \label{soln}%
    \end{figure}

Given the similar magnitudes of the early lightcurve and the
Mystery Spot (and indeed, the two are just phases of the same phenomenon),
the energetics are the same as posited in Meikle et al.~(1987),
except that the early lightcurve phase is shorter.  For an
interval of 10$^6$ s, at 5$\times$10$^{40}$ ergs s$^{-1}$, the optical
output, mostly from reprocessing of higher energy photons resulting
from the jet particles scattering with electrons, is 5$\times$10$^{46}$
ergs.  Since only a fraction of the particles scatter
in the polar ejecta, the overall efficiency, in the conversion of kinetic
energy into optical luminosity, could still be as low as
Meikle et al.'s estimated 0.001, which gives 5$\times$10$^{49}$
ergs of kinetic energy in the initial jet.  For 0.9 c protons,
each with 0.002 ergs of kinetic energy, this would mean 2.5$\times$10$^{52}$
protons or 2$\times$10$^{-5}$ M$_{\bigodot}$ initially each jet.
Without the now visible counterjet, the ``kick'' velocity to
the neutron star would be 10 km s$^{-1}$.  For a pulsar with
an initial spin rate of 500 Hz this short phase alone would
result in a drop of 10 Hz, corresponding to a mean spindown
rate of 10$^{-5}$ Hz s$^{-1}$, assuming a neutron star moment of 
inertia of 5$\times$10$^{44}$ gm-cm$^2$.

This may still be an underestimate, as accelerating a square 
$\ell$t-d of the polar ejecta column, which amounts to $\sim$0.002 M$_{\bigodot}$,
to $\sim$0.3 c, requires 1.6$\times$10$^{50}$ ergs of kinetic energy, 
which amounts to 6.6\% of the 2.5$\times$10$^{51}$ ergs of rotational energy 
of a 500 Hz pulsar for each jet, or $\sim$66 Hz of frequency 
drop from 500 Hz, still assuming 100\% conversion of jet kinetic energy 
into Mystery Spot kinetic energy, unless the plume has a smaller cross 
section than 1 $\ell$t-d$^2$, and/or the polar ejecta is less dense, on 
average, than 10$^7$.  Observations of initial pulsar spindowns (see 
Sect.~\ref{link}) would help greatly in resolving this
uncertainty.  Spinup from accretion may temper
the spindown somewhat (\cite{Pat09}), but gravitational
radiation reaction may counter it (\cite{Owen98}), though
the high electromagnetic spindown will mask any effect
of this latter on the pulsar braking index, $n$, where
${\partial f \over \partial t} \propto -f^{n}$, and
$n=5$ for pure gravitational radiation.

In either case, the rotational energy required is too large
to be supplied by a strongly magnetized pulsar over the required 
timescale, unless these are born spinning faster than the moderate 
rates generally believed to be typical (e.g., the 16.1 ms PSR 
J0537-6910 -- Middleditch et al.~2006).  There was certainly no 
evidence for a strongly magnetized pulsar within SN 1987A in its 
first few years (e.g., Pennypacker et al.~1989; \"Ogelmann et 
al.~1990; Kristian et al.~1991), and most importantly, there is 
no evidence for such a pulsar in the last few years (NASA et 
al.~2003), whereas SN 1986J, at the same age, showed clear evidence 
of such a
pulsar within it.\footnote{This SN, in the edge-on spiral
galaxy, NGC 0891, exceeds the luminosity of the Crab nebula at
15 GHz by a factor of 200 (\cite{Bie04}), and thus
is thought to have occurred because of a core collapse due to
iron photodissociation catastrophe
(FeSN), producing a {\it strongly} magnetized neutron star ($\sim$10$^{12}$
G -- the origin of magnetic fields in neutron stars is still poorly
understood, though it is believed that thermonuclear combustion
in a massive progenitor to an Fe core is related).}

However, there 
may be a weakly magnetized pulsar within SN 1987A (M00a,b), and 
at the very least this is supported by solid evidence for the 
formation of a neutron star (\cite{Bi87}; \cite{Hi87}).
A binary merger of two electron-degenerate stellar cores
(DD -- in isolation these would be white dwarfs) has been
proposed for SN 1987A (\cite{Pod89}), and the triple ring structure,
particularly that of the outer rings, has recently been successfully
calculated within this framework (Morris \& Podsiadlowski 2007).
Many other details of
87A, including the mixing (Fransson et al.~1989), the blue
supergiant progenitor (\cite{Sk69}), the early polarization (Schwarz \& Mundt
1987; Cropper et al.~1988; Barrett 1988), and the possible 2.14 ms optical 
pulsations (M00a,b), support this hypothesis.

The first clear evidence for DD-formed millisecond pulsars coincidentally 
came in the birth year of SN 1987A, with the discovery of the 3 ms pulsar, 
B1821-24 (\cite{Ly87}), in the non-core-collapsed globular cluster
M28.  Subsequently many more were found over the next 20 years in such 
clusters (e.g., 47 Tuc -- Camilo et al.~2000), and attributing these to 
recycling through
X-ray binaries has never really worked (\cite{CMR93}), by a few orders of
magnitude.\footnote{Relatively slowly rotating, recycled pulsars weighing
1.7 M$_\odot$, in the core-collapsed globular cluster, Ter 5 (\cite{Ra05}), 
have removed high
accretion rate from contention as a alternative mechanism to produce the
millisecond pulsars in the non-core-collapsed globular clusters.  The three 
millisecond pulsars in Ter 5 with periods $<$ 2 ms, Ter
5 O, P, and ad (\cite{He06}), may have been recycled starting with periods
near 2 ms.  There are four in this sample with periods between 2.05 and
2.24 ms, and perhaps most importantly, the first from Arecibo ALFA, 
1903+0327 (\cite{Ch08}), at 2.15 ms very close to the candidate 2.14 ms 
signature of SN 1987A (M00a,b), with a main-sequence companion, from
which it could never have accreted mass, nor significantly from any other 
source, because of its own modest mass.} 
Thus the DD process in SN 1987A, albeit within a common envelope, would 
likely have formed a rapidly spinning, weakly magnetized pulsar.

Consequently, we also argue, as a corollary implication useful for
understanding the SN process and its modern-day observation history, that
99\% of core-collapse events are similar to SN 1987A, in that they are
a result of the double-degenerate process, producing
only weakly magnetized, rapidly-spinning, millisecond pulsars, the notable
exceptions being SN 1986J and SN 2006gy, this latter which will be discussed 
further below.

\section{The SN 1987A link to GRBs}
\label{link}

Without the H and He in the envelope of the progenitor of 1987A
(or perhaps even with it),
Sk -69$^{\circ}$202, the collision of the jet with the 1987A polar 
ejecta (which produced the early light curve and Mystery Spot)
might be indistinguishable from a full $\ell$GRB
(\cite{Cen99}).\footnote{Otherwise it would just beg the question
of what distant, on-axis such objects would look like.}
This realization, together with the observation that
no $\ell$GRBs have been found in elliptical galaxies, and
the realization that the DD process {\it must} dominate (as
always, through binary-binary collisions), by a large factor the 
neutron star-neutron star mergers in these populations, even when 
requiring enough white dwarf-white dwarf merged
mass to produce core-collapse, leads to the unavoidable conclusion that
the DD process produces sGRBs in the absence of common envelope and 
polar ejecta, the means
by which they would otherwise become $\ell$GRBs.
Given that the sGRBs in ellipticals are due to mergers of white dwarfs,
we can conclude that:
1) the pre-common envelope/polar ejecta impact photon spectrum of 
$\ell$GRBs is well characterized,
2) sGRBs are offset from the centers of their elliptical hosts because
they are white dwarf-white dwarf mergers in their hosts' globular 
clusters (to produce most of their millisecond pulsars -- Gehrels et 
al.~2005), and 3) neutron star-neutron star mergers may not make GRBs 
as we know them, and/or be
as common as previously thought.\footnote{Thus
sGRBs may not flag neutron star-neutron star mergers, which may last only
a few ms, the same timescale as the 30-Jy, DM=375 radio burst
(\cite{Lo07}), far shorter than sGRBs (Hansen \& Lyutikov
2001).}

Thus Supernova 1987A, with its beam and jet producing its early 
light curve and MS, is potentially the Rosetta Stone for three of
the four types of GRBs, $\ell$, i, and s 
GRBs (\cite{Hor06}),\footnote{All except Soft Gamma Repeator
(SGR) GRBs, which are estimated to amount to less than 5\% of sGRBs
and 1.5\% of the total (\cite{Pa05}).}
with both polar ejecta and common envelope, red supergiant common 
envelope and no polar ejecta, and neither polar ejecta nor common
envelope, respectively (Middleditch 2007 -- hereafter M07).

In addition to axially driven pulsations, the SLIP model makes the
very unique and remarkable prediction that the component of pulsar
intensity which obeys Eqn.~1, diminishes only as distance$^{-1}$, 
and this has been verified experimentally (\cite{Ar04}), and also 
appears to be holding up (\cite{SSM09}), for pulsars in the Parkes 
Multibeam Survey (e.g., Lorimer et al.~2006).  There is also evidence 
that GRB afterglows share this characteristic (\cite{KK08}),
which supports the SLIP prediction of axially driven pulsations
when plasma is available at many $R_{LC}$.  The SLIP prediction
is convenient also because it explains how afterglows (and GRBs) 
can be visible across the Universe.\footnote{In the case of SN 1987A, 
the pulsations may have had to be observed through $\sim$13 $\ell$t-d
of polar ejecta.}
As a consequence of this prediction, we have initiated a campaign
of high speed monitoring of GRB afterglows.  

If, as for SN 1987A,
99\% of SNe are DD-initiated, then by measuring the pulse period,
$P$, of the optical/near infrared pulsations from an afterglow,
and assuming the pulsars resulting from DD are all produced
at a standard spin period, $P_0$, first measured from SN 1987A
near 2.14 ms, the redshift is given by:
\begin{equation}
z = {P \over P_0} - 1,
\end{equation}
and even a moderately precise $P$ (by standards), may yield a 
very precise redshift.

In the SLIP model, the peak of the emission for slowest pulsars occurs 
in the gamma-ray band (Ardavan et al.~2003,9), and this is supported by recent
gamma-ray detections of slow ($\sim$1 Hz) pulsars in supernova remnants 
by FERMI (e.g., Abdo et al.~2008).  There is no requirement in the SLIP
model on the rotator being a neutron star -- a white dwarf
will do as long as it has a magnetic field and there is
plasma outside of its light cylinder.  If this is the case,
strongly magnetized pulsars may not make GRBs, 
and it might even be possible for a pre-core-collapse,
$\sim$1.4 M$_{\bigodot}$ white dwarf, spinning at its minimum period of
1.98 s, to emit the prompt part of a GRB, and, as with the afterglow,
the distance$^{-1}$ law would likewise ameliorate the energy
requirement, even with the large expected spinup.  This also
raises the intriguing possibility that a GRB could be produced
without core-collapse, and a large spin-{\it down} may be present.  
We tested for spinup/down in the GRB with the highest fluence in 
the BATSE catalog, 960216 (\cite{Pa99}), by Fourier transforming the first
40 s of events and contouring power on the frequency-${\partial f \over
\partial t}$ plane.  Power appears, though not significant
without further confirmation, at a mean frequency of 0.50 Hz, and
derivative of +0.08 Hz s$^{-1}$, and also for spinup/down 
about an order of magnitude smaller, in the 0.35 to 0.42 Hz region. 
Bursts with even better statistics (perhaps from FERMI)
may be necessary to further test this hypothesis.

The geometric model with small angle scattering of gamma-rays in
distant polar ejecta can explain other details of $\ell$GRBs,
such as their $\sim$100 s T$_{90}$'s,\footnote{An offset of 
0.5$^{\circ}$ at $\sim$10 $\ell$t-d corresponds to a 33 s delay.}
the negligible spectral lag for late ($\sim$10--100 s) emission
from ``spikelike'' bursts (\cite{NB06}), and why ``precursor'' and
``delayed'' contributions have similar temporal structure
(Nakar \& Piran 2002; M07).  

\section{Double-Degenerate in Type Ia/c SNe}
\label{Ia/c}

Since 2007, Feb., it appeared unavoidable that Type Ia SNe were
also DD-caused, one of the causes being the long list of reasons
why Ia's can not be due to thermonuclear disruption (M07).  Now it 
is not clear if this ever happens in {\it any} progenitor (see, e.g., 
Seigfried 2007), and empty SN remnants
almost always contain a neutron star which is just not visible, just
as the one in Cas A is barely visible (\cite{Ta99}).  Further, this means
that Ia's and Ic's (these latter have been regarded by many as DD-initiated,
neutron star-producing since the invention of the classification), are both 
due to the DD process, and thus must form a continuous class: Ia's when viewed
from the merger equator, with lines of Fe; and Ic's when viewed from the
merger poles, because this view reveals lines of the r-process
elements characteristic of Ic's,\footnote{If sufficient matter exists,
in excess of that lost to core-collapse, to screen the Ia thermonuclear 
products -- a rare circumstance in elliptical galaxies, the reason why
Ic's are absent from these.}
part of the reason for the differing spectroscopic classification.
The high approaching
velocities frequently seen in Ic's (e.g., ``hypernovae'') are due to 
the view looking down the the axis of the approaching bipolarity.

In the application of Type Ia SN luminosities for cosmological
purposes, the increase in blue magnitude from the light curve maximum, 
$\Delta M_B$ (essentially an inverse measure of the width of the
light curve in time), measured in the first few weeks of SN Ia proper 
time, is used to correct the Ia luminosity for the variable amount of 
$^{56}$Ni produced (\cite{Ph93}).  
However, the direct relation, between the $\Delta M _B$ of the
width-luminosity relation and the fractional SiII polarization in Ia's,
pointed out by Wang et al.~(2006), is more meaningfully interpreted
as an {\it inverse} relation between the SiII polarization and
luminosity (unlike the Fe lines, SiII lines must also exist in the
axial features because they are also observed in Ic's, and their
polarization in Ia's is a result of the more rapid axial
extension when viewed close to the merger equator).
This inverse relation would be expected in Ia's if the luminosity 
of the axial features were fixed, while the luminosity of the toroidal
component is driven by the amount of encapsulated $^{56}$Ni
positron annihilation gamma-rays, which can be very
high.\footnote{As with 1987A-like events, it
would again beg the question of ``What {\it else} they could possibly
be?,'' and ``delayed detonation'' (\cite{Kh91}), or
``gravitationally confined detonation'' (\cite{Pl04}), do not produce
polarization which would be inversely proportional to luminosity.  And 
unless the view {\it is} very near polar,
this geometry can produce split emission
line(s) on rare occasions, as was seen in SN 2003jd (\cite{Maz05b}),
and thus again there is no need to invoke exotica, or an entire
population (III) to account for GRBs (Conselice et al.~2005; M04).}

Because there is a spectroscopic difference between Ia's and Ic's,
the SLIP-driven polar jets are either deficient in $^{56}$Ni, or
are too diffuse to encapsulate their gamma-rays,
or both.  {\it No} observation of {\it any} recent SN other than SN 1986J
and SN 2006gy, including all {\it ever} made of Type Ia SNe, is
inconsistent with the bipolar geometry of 1987A.

All this raises serious concerns about the use of SNe Ia in cosmology, 
because many Ia/c's in actively star-forming galaxies belong to the continuous 
class, and some of these, and most Ia's in ellipticals, may not
encapsulate a sufficient fraction of their gamma-rays to be bolometric
(\cite{PE01}), especially given the toroidal geometry, lying as much as two 
whole magnitudes below the width-luminosity relation (the faint SNe Ia of 
Benetti et al.~2005).  In the SLIP model the pulsar eviscerates its stellar 
remnant as long as there is
remnant remaining, enforcing a toroidal geometry of ejecta near
its rotation equator.  Even if this toroid is very much brighter
than the axial jets, as is the case in many Ia/c's, the {\it opacity}
of the axial jets, in front of rear sections of the toroid, which would
otherwise be visible even for small inclinations away from 90$^{\circ}$ 
(a viewing angle for which essentially all SNe Ia/c will be classified as 
Ia's), may change during the interval when the width-luminosity relation 
is measured, literally and figuratively casting a shadow of reasonable 
doubt over attempts to use Ia/c's as cosmological standard candles.

\section{Type II SNe and Other Details}
\label{TypeII}

The double-degenerate mechanism ensures that nearly all SNe are born 
from a post-merger white dwarf with a rotation period near 1.98 s, thus rapid
rotation can not be invoked as an unusual
circumstance, for the case of SN 2003fg, to justify
``super-Chandrasekhar-mass'' white dwarfs.  The $>$1.2 M$_{\odot}$
of $^{56}$Ni it produced may only mean that core collapse underneath 
mixed thermonuclear fuel can initiate very efficient
combustion/detonation,\footnote{The spectroscopic
demands of a significant mass of unburned fuel, such as O, being
invalid because of the invalid paradigm under which such
estimates were made.}
the paltry amounts
of $^{56}$Ni associated with Ib's and at least 90\% of
IIs being the result of dilution of their thermonuclear fuel with He
and/or H due to the DD merger process.\footnote{Helium has
been found where it was not expected in almost all well-studied
SNe.}
Thus SN 2006gy (\cite{Sm07}) may not
be a pair-instability SN (\cite{Wo07}; \cite{Ka09}), or a collision of two
massive stars (\cite{PZ07}),\footnote{The inner
layers of all FeSNe, possibly {\it many} M$_{\odot}$ of Si,
Ne, O, and C, have not been diluted with He by DD, and thus
may ignite/detonate upon core collapse, and burn efficiently.  SN 
modelers therefore face the unenviable choice of calculating FeSNe,
which involve strong magnetic fields, or DD SNe, which involve a
great deal of angular momentum, and {\it demand} GRBs as an
outcome (see Sect.~\ref{link}).}
only a massive FeSN of up to 75 M$_{\bigodot}$, which may
actually have produced several M$_{\odot}$
of $^{56}$Ni, {\it and} a strongly magnetized neutron star remnant, a
prediction which can be tested soon.\footnote{As a corollary,
40 M$_{\bigodot}$ in a SN remnant is no longer a reason
to invoke ``millisecond magnetars,'' as the dispersal
mechanism (\cite{Th04},\cite{JV06})}.

The presence of plasma makes a huge difference to rapidly
rotating, weakly magnetized neutron stars.  Strong pulsations
have occurred during observations of SN remnants or X-ray
binaries which have never been subsequently confirmed, and yet
have no explanation other than as a real, astrophysical source
(e.g., Bleach et al.~1975).  Judging from the high fraction
of empty SN remnants, the population of ``quiet'' neutron
stars (\cite{Wea94}) must exceed all other ``loud'' populations
combined.  Only when such a weakly-magnetized, rapidly-spinning
neutron star encounters a cloud of matter will it become
sufficiently luminous to be detected.  In the context of the SLIP 
model, radiation from the known millisecond pulsars may very
well be detected from Earth because we are ``in the cusp,''
i.e., we are in the part of the pulsar's beam (Eqn.~1) that
decays inversely only as the first power of distance.
If this is not the case, young neutron stars may only
appear as thermal sources, such as the one in Cas A
(\cite{Ta99}).  A century older, as is the case with
Tycho 1572, or more deeply embedded in the Galactic
plane, as is the case for Kepler, 1604, and not even
the thermal sources are detectable.  Evidence does
linger, however, at least in the outwardly very sphericial 
remnant of SN 1006, as bipolar high energy emission in XMM 
and VLA images (\cite{Ro04}).

In the case of SN 1987A, plasma initially available at many 
$R_{LC}$ resulted in axially driven pulsations.  As 
the circum-neutron star density declined, polarization
currents were restricted to fewer $R_{LC}$, resulting in 
pulsed emission along a cone of finite polar angle, which 
may have modified the resulting beam/jet 
into the approaching and retreating conical features now 
easily visible in the HST ACS images (\cite{Ki03}).  Eventually, 
as the plasma continued to thin with time, its maximum density 
occurred between 2$R_{LC}$ and just outside of $R_{LC}$, 
resulting in pulsations driven close the pulsar's rotational 
equator, and according to our self-consistent solution, in the line 
of sight to the Earth.  Precession and nutation may have further 
embellished the axial pattern (M00a).  Even if totally absorbed,
such a beam would produce an observable excess luminosity that may
have been seen by 1991 (Cowan 1991; Suntzeff et al.~1991,2), as
the amount of $^{57}$Co required to otherwise account for
the excess was only barely consistent with hard X-ray
and infrared spectral data (\cite{Kum89}; \cite{Rank88}).  A few
years earlier it is unlikely that the 2.14 ms signal would have
been detectable in the broadband, as limits established in early
1988 (\cite{Pe89}) are comparable to levels of the 2.14 ms signal
observed in the I band between 1992, Feb.~and 1993, Feb.  The
2.14 ms pulsar candidate was last detected in 1996,
Feb. (M00b), and by 2002 there was no evidence of
any such source in ACS images, which only really means
that any pulsar within SN 1987A had entered the
``Cas A'' phase,\footnote{Calculations with the SLIP model 
involve a variation of Kepler's equation, which relates the 
eccentric anomaly, $E$, to the mean anomaly, $M$, using the 
eccentricity, $\epsilon$, $\rm{E} - \epsilon \sin \rm{E} = \rm{M}$,
but in this case $\epsilon > 1$.  Such calculations are notoriously
difficult, even for a compact star {\it not} surrounded
by remnant plasma.  Needless to say, no such calculations
have been done to date, and thus no calculation done so
far, including those of ``collapsars,'' can possibly be
valid.  One side effect of not properly accounting for the
pulsar, and the large amount of $^{56}$Ni when strongly-magnetized
pulsars are produced, is a very low estimate for the mass,
$\sim$25 M$_{\bigodot}$, above which the core collapse continues on
to a black hole.  SN 2006gy, with several M$_{\bigodot}$
of $^{56}$Ni, exposed this delusion.}
having exhausted its surrounding plasma
supply and perhaps also because the Earth was no longer
in the ``cusp'' of its beam(s).  Still, the central
source should turn on when the pulsar encounters matter
from time to time.

A beam of protons, with kinetic energies of up to 2.2 GeV or 
greater, will eventually produce electrons with similar energies.  
Even higher energies may
result from core-collapse events with less material in the
common envelope, and/or these may, in turn,
be further accelerated by magnetic reconnection, in wound-up
magnetic fields near the Galactic center, or other mechanisms
(\cite{Sch09}), possibly to TeV
energies, to produce the WMAP ``haze'' observed in that
direction (\cite{Fink04}).  In addition, the loss of positrons,
which occurs because of the bipolarity of SNe, which also makes 
them unfit for easy cosmological interpretation, may show up as
an excess in cosmic ray data (\cite{Chang08}; \cite{Ab09}; \cite{Adri09}),
a satisfying resolution for the apparent anomalous dimming of
distant SNe explained in terms of local cosmic ray abundances.

\section{Conclusion}
\label{conc}

We have derived a self-consistent solution for the onset
($\sim$11 $\ell$t-d), and depth ($\sim$13 $\ell$t-d),
of the polar ejecta of the progenitor of SN 1987A,
the energetics of its beam/enhanced UV flash, the kinetics
of its jet, and angle from the line of sight to
the Earth ($\sim$75$^{\circ}$).  There is plenty of
evidence for the absence of any strongly magnetized pulsar
within SN 1987A, and such a pulsar may not have the rotational 
energy to account for the kinetics anyway.
Thus, we have argued, through the paradigm of a model of pulsar
emission from superluminally induced polarization currents (SLIP)
which uses emission from polarization currents
induced beyond the pulsar light cylinder (\cite{Ar98}),
that SN 1987A, its beam/jet, ``Mystery Spot,'' and possible
2.14 ms pulsar remnant, are intimately related to as many as
99\% of GRBs, millisecond pulsars, and other SNe, including all 
Type Ia SNe.  The SLIP model explains, in a natural way (Eqn.~1), 
the changes over time observed in the collimation of the SN 1987A 
beam and jet.

The time lags, energetics, and collimation of $\ell$GRBs are
consistent with those of 1987A's beam, jet, and ``Mystery Spot''.
When the bipolarity of SN 1987A is interpreted through this 
model, its pulsar clearly had ablated the $\sim$10 M$_{\bigodot}$ 
of ejecta, eviscerating the remnant by blowing 
matter out of its poles at speeds up to 0.95 c or greater, 
and enforcing a toroidal geometry on the remaining
equatorial ejecta.  Since there is no reason to suggest that
this is not universally applicable to all SNe, this geometry
has grave implications for the use of Type Ia SNe as standard
candles in cosmology.

The interaction of even a weakly magnetized pulsar with the rest
of the remnant of the progenitor (if this rest exists), clearly
can not be ignored.  There appears to be no need to invent exotica
to explain GRBs -- the SLIP model provides the young pulsar (or even
a near-Chandrasekhar-mass white dwarf) as the central engine, and makes 
the very specific and testable prediction that GRB afterglows are, 
in fact, pulsars.  In addition, because of the unsuitability of SN 
geometries for cosmological interpretation, the expansion of the 
Universe may not be accelerating, and, as a consequence, there may 
be no dark energy.  But if there is no dark energy, then there
is no numerical coincidence to support the role of dark
matter in Concordance Cosmology.  Recent observations have
also cast significant doubt on the existence of dark matter
(\cite{NP07}; \cite{MNP09}).

Although it might appear that a Universe without dark matter
or energy, collapsars, pair instability SNe, super-Chandrasekhar 
mass white dwarfs, frequent collisions of massive stars,
and neutron star-neutron star mergers which make sGRBs, is much less
``exotic'' than previously thought, pulsars, i.e., clocks and
minutes-old neutron stars to boot, which can be seen across
almost the entire Universe, may well suffice in explaining all
of the issues which gave rise to the previously mentioned 
entities, are more in line with Occam's Razor, and are also, 
of themselves, extremely worthy of study.

\begin{acknowledgements}
I would like to thank Dr.~John Singleton for supporting this
work through LDRD grant DR20080085, as well as Drs.~Joe Fasel,
Bill Junor, and Andrea Schmidt.  This work was performed
under the auspices of the Department of Energy.
\end{acknowledgements}


\begin{thebibliography}{}
\bibitem[Abdo et al.\ 2008]{Ab08} Abdo, A. A., Ackerman, M., Atwood, W. B., 
    et al. 2008, Science, 322, 1218
\bibitem[Abdo et al.\ 2009]{Ab09} Abdo, A. A.,  Ackerman, M., Ajello, M., 
    et al. 2009, Phys. Rev.  Lett., 102, 1101
\bibitem[Adriani et al.\ 2009]{Adri09} Adriani, O., Barbarino, G. C., 
    Bazilevskaya, G. A., et al. 2009, \nat, 458, 607
\bibitem[Ardavan 1994]{Ar94} Ardavan, H. 1994, \mnras, 268, 361
\bibitem[Ardavan 1998]{Ar98} Ardavan, H. 1998, Phys. Rev. E., 58, 6659
\bibitem[Ardavan et al.\ 2003]{Ar03} Ardavan, H., Ardavan, A., \& Singleton, J.
    2003, JOSAA, 20, 2137
\bibitem[Ardavan et al.\ 2004]{Ar04} Ardavan, A., Hayes, W., Singleton, J.,
    et al. 2004, J. Appl. Phys., 96, 4614
\bibitem[Ardavan et al.\ 2008]{Ar08} Ardavan, H., Ardavan, A., Singleton, J.,
    \& Perez, M. R. 2008, \mnras, 388, 873
\bibitem[Ardavan et al.\ 2009]{Ar09} Ardavan, H., Ardavan, A., Singleton, J.,
    et al. 2009, \mnras, in preparation
\bibitem[Barrett 1988]{Ba88} Barrett, P. 1988, \mnras, 234, 937
\bibitem[Benetti et al.\ 2005]{Be05} Benetti, S., Capillaro, E., Mazzali, P. A., 
    et al. 2005, \apj, 623, 1011
\bibitem[Bietenholz et al.\ 2004]{Bie04} Bietenholz, M. F.,
    Bartel, N., \& Rupen, M. P. 2004, Science, 304, 1947
\bibitem[Bionta et al.\ 1987]{Bi87} Bionta, R. M., Blewit, G., Bratton, C. B.,
    Caspere, D., \& Ciocio, A. 1987, Phys. Rev. Lett., 58, 1494
\bibitem[Bleach et al.\ 1975]{Bl75} Bleach, R. D., Henry, R. C., Fritz, G.,
    et al. 1975, \apj, 197, L13
\bibitem[Braje \& Romani 2001]{BR01} Braje, T. M., \& Romani, R. 2001, \apj, 550,
    392
\bibitem[Camilo et al.\ 2000]{Ca00} Camilo, F., Lorimer, D. R., Freire, P.,
    Lyne, A. G., \& Manchester, R. N. 2000, \apj, 535, 975
\bibitem[Cen 1999]{Cen99} Cen, R. 1999, \apjl, 524, L51
\bibitem[Champion et al.\ 2008]{Ch08} Champion, D. J., Ransom, S. M., Lazarus, P.,
    et al. 2008, Science, 320, 1309
\bibitem[Chang et al.\ 2008]{Chang08}Chang, J., Adams, J. H., Ahn, H. S.,
    et al. 2008, \nat, 456, 362
\bibitem[Chen et al.\ 1993]{CMR93} Chen, K., Middleditch, J., \& Ruderman, M. 1993,
    \apj, 408, L17
\bibitem[Conselice et al.\ 2005]{Con05} Conselice, C. J., Vreeswijk, P. M.,
    Fruchter, A. A., et al. 2005, \apj, 633, 29
\bibitem[Cowan 1991]{Cow91} Cowan, R. 1991, Science News, October 19, 1991
\bibitem[Cropper et al.\ 1988]{Cr88} Cropper, M., Bailey, J., McCowage, J., et al. 1988,
    \mnras, 231, 695
\bibitem[Danziger et al.\ 1987]{Da87} Danziger, I. J., Fosbury, R. A. E., Alloin, D., 
    et al. 1987, \aap, 177, L13
\bibitem[Finkbeiner 2004]{Fink04}Finkbeiner, D. P. 2004, \apj, 614, 186
\bibitem[Fransson et al.\ 1989]{Fr89}Fransson, C., Cassatella, A., Gilmozzi, R., 
    et al. 1989, \apj, 336, 429.
\bibitem[Gehrels et al.\ 2005]{Gh5} Gehrels, N., Sarazin, C. L., O'Brien, P.  T., 
    et al. 2005, \nat, 437, 851
\bibitem[Hamuy \& Suntzeff 1990]{HS90} Hamuy, M., \& Suntzeff, N. B. 1990, \aj,
    99, 1146
\bibitem[Hansen \& Lyutikov 2001]{Ha01} Hansen, B. M. S., \& Lyutikov, M. 2001,
    \mnras, 322, 695
\bibitem[Hessels et al.\ 2006]{He06} Hessels, J. W. T., Ransom, S. M., Stairs, I.
    H., et al. 2006, Science, 311, 1901
\bibitem[Hester et al.\ 2002]{He02} Hester, J. J., Mori, K., Burrows, D.,
    et al. 2002, \apj, 577, L49
\bibitem[Hirata et al.\ 1987]{Hi87} Hirata, K., Kajita, T., Koshiba, M.,
    Nakahata, M., \& Oyama, M. 1987, Phys. Rev. Lett., 58, 1490
\bibitem[Horv\'ath et al.\ 2006]{Hor06} Horv\'ath, I., Bal\'azs, L. G., Bagoly,
    Z., Ryde, F., \& A M\'ez\'aros, A. 2006, \aap, 447, 23
\bibitem[Kann \& Klose 2008]{KK08}Kann, D. A., \& Klose S. 2008, in Supernova
    1987A: 20 Years After, Supernovae and Gamma-Ray Bursters, ed. S. Immler,
    K. W. Weiler, \& R. McCray (New York:AIP) 107, 293
\bibitem[Kawabata et al.\ 2009]{Ka09}Kawabata, K. J., Tanaka, M., Maeda, K., et al.
    2009, \apj, 697, 747
\bibitem[Khokhlov 1991]{Kh91} Khokhlov, A. M. 1991, \aap, 245, 114
\bibitem[Kobayashi et al.\ 2006]{Ko06} Kobayashi, C., Umeda, H., Nomoto, K.,
    Tominaga, N., \& Ohkubo, T. 2006, \apj, 653 1145
\bibitem[Kristian et al.\ 1991]{Kr91} Kristian, J. A., Pennypacker, C. R., 
    Middleditch, J., et al. 1991, \nat, 349, 747
\bibitem[Kumagai et al.\ 1989]{Kum89} Kumagai, S., Shigeyama, T., Nomoto, K.,
    et al. 1989, \apj, 345, 412
\bibitem[Kuo-Petravic et al.\ 1974]{KPR74} Kuo-Petravic, L. G., Petravic, M.,
    \& Roberts, K. V. 1974, \prl, 32, 18, 1019
\bibitem[Lorimer et al.\ 2006]{LFL06} Lorimer, D. R., Faulkner, A. J., Lyne, A. G., 
    et al. 2006, \mnras, 372, 777
\bibitem[Lorimer et al.\ 2007]{Lo07} Lorimer, D. R., Bailes, M., McLaughlin, M. A.,
    Narkevic, D. J., \& Crawford, F. 2007, Science, 318, 777
\bibitem[Lyne et al.\ 1987]{Ly87} Lyne, A. G., Brinklow, A., Middleditch, J.,
    Kulkarni, S. R., \& Backer, D. C. 1987, \nat, 328, 399
\bibitem[MacFadyen \& Woosley 1999]{MW99} MacFadyen, A. I., \& Woosley, S. E. 1999,
    \apj, 524, 262
\bibitem[Madore et al.\ 2009]{MNP09} Madore, B. F., Nelson, E., \& Petrillo, K.
    2009, \apjs, 181, 572
\bibitem[Mazzali et al.\ 2005]{Maz05b} Mazzali, P., Kawabata, K. S., Maeda, K., 
    et al. 2005, Science, 308, 1284
\bibitem[Meikle et al.\ 1987]{Me87} Meikle, W. P. S., Matcher, S. J., \& Morgan, B. L.
    1987, \nat, 329, 608
\bibitem[Menzies et al.\ 1987]{Men87} Menzies, J. W., Catchpole, R. M., van Vuuren, K., 
    et al. 1987, \mnras, 227, 39P
\bibitem[M\'esz\'aros 2006]{Mz06}M\'esz\'aros, P. 2006, Rep. Prog. Phys., 69,
    2259
\bibitem[Middleditch et al.\ 2000a]{M00a}Middleditch, J., Kristian, J. A., Kunkel 
    W. E., et al. 2000a, \na, 5, 243 (M00a)
\bibitem[Middleditch et al.\ 2000b]{M00b}Middleditch, J., Kristian, J. A., Kunkel, 
    W. E., et al. 2000b, preprint (astro-ph/0010044) (M00b)
\bibitem[Middleditch 2004]{M04}Middleditch, J. 2004, \apjl, 601, L167 (M04)
\bibitem[Middleditch et al.\ 2006]{M06}Middleditch, J., Marshall, F. E., Wang, Q. D.,
    Gotthelf, E. V., \& Zhang, W. 2006, \apj, 652, 1531
\bibitem[Middleditch 2007]{M07}Middleditch, J. 2007, preprint, ArXiv/0705.2263 (M07)
\bibitem[Middleditch \& Perez 2008]{M08}Middleditch, J., \& Perez, M. R. 2008, BAAS, 40, 206,
    http://www.ccs3.lanl.gov/\~{}jon/pdf/stl2.pdf
\bibitem[Morris \& Podsiadlowski 2007]{MP07} Morris, T., \& Podsiadlowski, Ph. 2007,
    Science, 315, 1103
\bibitem[Nakar \& Piran 2002]{NP02} Nakar, E., \& Piran, T. 2002, \mnras, 331, 40
\bibitem[NASA et al.\ 2003]{Ki03} NASA, Challis, P., Kirshner, R. P., \& Sugerman, B.
    2003, http://hubblesite.org/gallery/album/entire\_collection/pr2004009a/
    http://hubblesite.org/gallery/album/entire\_collection/pr2007010a.
\bibitem[Nelson \& Petrillo 2007]{NP07}Nelson, E., \& Petrillo, K. 2007, \baas, 39,
    1, 184
\bibitem[Nisenson et al.\ 1987]{Ni87}Nisenson, P., Papaliolios, C., Karovska, M.,
    \& Noyes, R. 1987, \apjl, 320, L15
\bibitem[Norris \& Bonnell 2006]{NB06} Norris, J. P., \& Bonnell, J. T. 2006, \apj,
    643, 266
\bibitem[\"Ogelmann et al.\ 1990]{Og90}\"Ogelmann, H., Gouiffes, C., Augusteijn, 
    T., et al. 1990, \aap, 237, L9
\bibitem[Owen et al.\ 1998]{Owen98}Owen, B., Lindblom, L., Cutler, C., et al.  1998, 
    Phys. Rev. D., 58, 084020
\bibitem[Paciesas et al.\ 1999]{Pa99} Paciesas, W. S., Meegan, C. A., Pendleton, 
    G. N., et al. 1999, \apjs, 122, 465
\bibitem[Palmer et al.\ 2005]{Pa05} Palmer, D., Barthelmy, S., Gehrels, N.,
    et al. 2005, \nat, 434, 1107
\bibitem[Panagia et al.\ 2006]{Pa06} Panagia, N., Van Dyk, S. D., Weiler, K. W.,
    et al. 2006, \apj, 646, 959
\bibitem[Patruno et al.\ 2009]{Pat09} Patruno, A., Hartman, J. M., Wijnands, R.,
    Chakrabarty, D., \& van der Klis, M. 2009, preprint, ArXiv:09024323
\bibitem[Pennypacker et al.\ 1989]{Pe89} Pennypacker, C. R., Kristian, J. A.,
    Middleditch, J, et al. 1989, \apj, 340, L61
\bibitem[Phillips 1993]{Ph93} Phillips, M. M. 1993, \apj, 413, L105
\bibitem[Pinto \& Eastman 2001]{PE01} Pinto, P. A., \& Eastman, R. G. 2001, \na, 6,
    307
\bibitem[Piran \& Nakamura 1987]{PN87} Piran, T., \& Nakamura, T. 1987, \nat, 330,
    28
\bibitem[Plewa et al.\ 2004]{Pl04} Plewa, T., Calder, A. C., \& Lamb, D. Q. 2004,
    \apjl, 612, L37
\bibitem[Podsiadlowski \& Joss 1989]{Pod89}Podsiadlowski, Ph., \& Joss, P.~C.
    1989, \nat, 338, 401
\bibitem[Portegies Zwart \& van den Heuvel 2007]{PZ07}Portegies Zwart, S. F., \& van den 
    Heuvel, E. P. J. 2007, \nat, 450, 388
\bibitem[Rank et al.\ 1988]{Rank88} Rank, D. M., Pinto, P. A., Woosley, S. E., Bregman,
    J. D., \& Witteborn, F. C. 1988, \nat, 331, 505
\bibitem[Ransom et al.\ 2005]{Ra05} Ransom, S. M., Hessels, J. W. T., Stairs, I. H.
    et al. 2005, Science, 307, 892
\bibitem[Rees 1987]{Rees87}Rees, M. 1987, \nat, 328, 207
\bibitem[Rothenflug et al.\ 2004]{Ro04} Rothenflug, R., Ballet, J., Dubner,
    G., et al. 2004, \aap, 425, 121
\bibitem[Sanduleak 1969]{Sk69} Sanduleak, N. 1969, Contr. CTIO, 1969
\bibitem[Schure et al.\ 2009]{Sch09} Schure, K. M., Vink, J., Achterberg, A., 
    \& Keppens, R. 2009, Adv. Sp. Res., 44, 433
\bibitem[Schwarz \& Mundt 1987]{Sc87} Schwarz, H. E., \& Mundt, R. 1987, \aap, 177, L4
\bibitem[Siegfried 2007]{Sieg07} Siegfried, T. 2007, Science, 316, 194
\bibitem[Singleton et al. 2009]{SSM09} Singleton, J., Sengupta, P., Middleditch,
    J., et al. 2009, \nat, submitted
\bibitem[Smith et al.\ 2007]{Sm07}Smith, N., Li, W., Foley, R. J., et al. 2007, 
    \apj, 666, 1116
\bibitem[Sugerman et al.\ 2005]{Su05} Sugerman, B. E. K., Crotts, A. P. S.,
    Kunkel, W. E., Heathcote, S. R., \& Lawrence, S. S. 2005, \apj, 627, 888
\bibitem[Suntzeff et al.\ 1991]{Sun91} Suntzeff, N., B., Phillips, M. M., Elias,
    J. H., Depoy, D. L., \& Walker, A. R. 1991, \aj, 102, 1118
\bibitem[Suntzeff et al.\ 1992]{Sun92} Suntzeff, N., B., Phillips, M. M., Elias,
    J. H., Walker, A. R., \& Depoy, D. L. 1992, \apjl, 384, 33
\bibitem[Tananbaum et al.\ 1999]{Ta99} Tananbaum, H., \& The Chandra Observing
    Team 1999, \iaucirc, No. 7246, 1
\bibitem[Thompson et al.\ 2004]{Th04} Thompson, T. A., Chang, P., \& Quataert,
    E. 2004, \apj, 611, 380
\bibitem[Vink \& Kuiper 2006]{JV06} Vink, J., \& Kuiper, L. 2006, \mnras, 370, L14
\bibitem[Wamsteker et al.\ 1987]{Wa87} Wamsteker, W., Panagia, N., Barylak, M., 
    et al. 1987, \aap, 177, L21
\bibitem[Wang et al.\ 2002]{Wa02} Wang, L., Wheeler, J. C., Hoflich, P., 
    et al. 2002, \apj, 579, 671
\bibitem[Wang et al.\ 2006]{Wa06} Wang, L., Baade, D., \& Patat, P. 2006, Science,
    315, 212
\bibitem[Weatherall 1994]{Wea94} Weatherall, J. 1994, \apj, 428, 261
\bibitem[Woosley et al.\ 2007]{Wo07} Woosley, S. E., Blinnikov, S., \& Heger, A.
    2007, \nat, 450, 390
\end{thebibliography}
\end{document}